\documentclass{vgtc}                          

\ifpdf
  \pdfoutput=1\relax                   
  \usepackage{graphicx}                
  \DeclareGraphicsExtensions{.pdf,.png,.jpg,.jpeg} 
\else
  \usepackage{graphicx}                
  \DeclareGraphicsExtensions{.eps}     
\fi%

\graphicspath{{figures/}{pictures/}{images/}{./}} 

\usepackage{microtype}                 
\PassOptionsToPackage{warn}{textcomp}  
\usepackage{textcomp}                  
\usepackage{mathptmx}                  
\usepackage{times}                     
\usepackage{cite}                      
\usepackage{tabu}                      
\usepackage{booktabs}                  

\onlineid{0}

\vgtccategory{Research}

\vgtcinsertpkg

\title{PanoCoach: Enhancing Tactical Coaching and Communication in Soccer with Mixed-Reality Telepresence}

\author{Andrew Kang\thanks{e-mail: ak108@rice.edu. Work done during an internship at Harvard.}\\ %
        \scriptsize Harvard University \\ \scriptsize Rice University%
\and Hanspeter Pfister\thanks{e-mail: pfister@g.harvard.edu}\\ %
     \scriptsize Harvard University
     \and Tica Lin\thanks{e-mail: mlin@g.harvard.edu}\\ %
     \scriptsize Harvard University%
     }

\teaser{
  \includegraphics[width=\linewidth, alt={}]{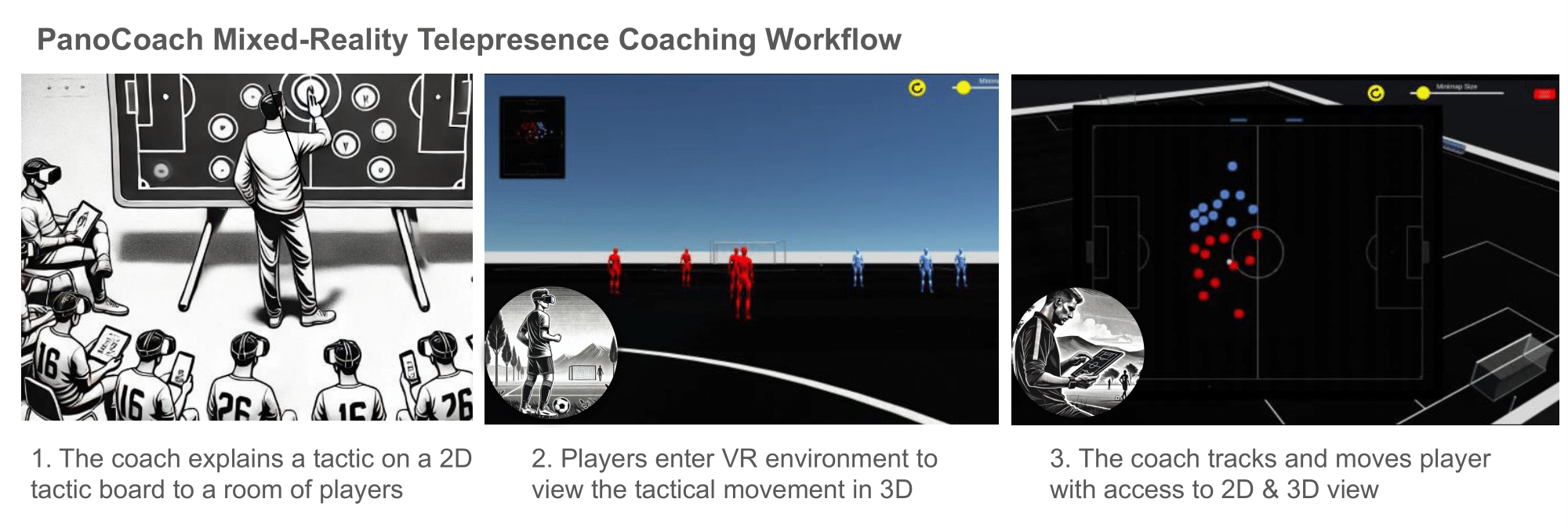}
  \caption{%
  	\toolName{} enhances tactical coaching by synchronizing a 2D tablet view with a simulated 3D environment. The proposed mixed-reality telepresence approach enables players to develop spatial understanding of complex team tactics from an immersive first-person perspective, while allowing coaches to maintain the ability to annotate and reposition players on the 2D board. This setup ensures that players receive immediate feedback from the coach while engaged in the 3D environment. Conceptual illustrations are generated by Dall-E. \toolName{} prototype views were implemented with Unity. 
  }
  \label{fig:teaser}
}

\begin{document}

\newcommand{\toolName}{PanoCoach}


\firstsection{Introduction}

\maketitle

\firstsection{Introduction} 
Soccer, as a dynamic team sport, requires seamless coordination and integration of tactical strategies across all players. Adapting to new tactical systems is a critical but often challenging aspect of soccer at all professional levels. Even the best players can struggle with this process, primarily due to the complexities of conveying and internalizing intricate tactical patterns. Traditional communication methods like whiteboards, on-field instructions, and video analysis often present significant difficulties in perceiving spatial relationships, anticipating team movements, and facilitating live conversation during training sessions. These challenges can lead to inconsistent interpretations of the coach’s tactics among players, regardless of their skill level.

To bridge the gap between tactical communication and physical execution, we propose a mixed-reality telepresence solution, \toolName{}, designed to support multi-view tactical explanations during practice. Our concept involves a multi-screen setup combining a tablet for coaches to annotate and demonstrate concepts in both 2D and 3D views, alongside VR to immerse athletes in a first-person perspective, allowing them to experience a sense of presence during coaching.

In our preliminary study, we prototyped the cross-device functionality to implement the key steps of our approach: Step 1, where the coach uses a tablet to provide clear and dynamic tactical instructions, Step 2, where players engage with these instructions through an immersive VR experience, and Step 3, where the coach tracks players' movements and provides real time feedback. 
User evaluation with coaches at City Football Group, Harvard Soccer and Rice Soccer suggests this mixed-reality telepresence approach holds promising potential for improving tactical understanding and communication.

Based on these findings, we outline future directions and discuss the research needed to expand this approach beyond controlled indoor environments, such as locker rooms, leveraging telepresence to enhance tactical comprehension and simulated training.

\section{Related Work}

The potential of VR to assess and train team sports performance has been explored extensively. 
Faure et al. \cite{faure2019} highlight VR's capability to simulate realistic game environments for training in team ball sports. Studies have demonstrated that visual stimuli training programs can significantly enhance cognitive function, 
reaction time, and spatial awareness in soccer players \cite{theofilou2022}. Fortes et al. \cite{fortes2021} argue that VR offers more effective 
perceptual-cognitive skill development compared to traditional video-stimulation training, making it particularly 
beneficial for young athletes. More generally, VR has proven advantageous for enhancing anticipatory 
performance and perception-action coupling in sports like baseball and rugby \cite{ranganathan2007, correia2012}. Van Maarseveen et al. \cite{vanMaarseveen2018} 
and Theofilou et al. \cite{theofilou2022} emphasize the importance of perceptual-cognitive skills in situ, where players must 
make decisions based on real-time affordances within dynamic environments. 

VR Applications are also being extended to sports other than soccer, such as basketball\cite{tsai2022},  baseball\cite{liu2024} and badminton~\cite{lin2024},
showing the positive impact of VR-based training on tactical awareness, reaction time, and even batting
performance through AI-driven feedback systems. In dance and motion-based activities, VR systems like
WAVE\cite{laattala2024} demonstrate how anticipatory movement visualization can be adapted for training, an approach that has
potential crossover benefits in high-performance sports such as soccer. Furthermore, animated VR and 360 degree
VR are becoming prominent tools for the evaluation and training of decision making in team sports, offering unique
insights into spatial relationships and tactical understanding \cite{jia2024}.
 
While previous research has focused on enhancing motion learning and decision-making with VR, few studies have addressed facilitating real-time coaching and communication, particularly for complex team tactics that require coordinated teamwork. Our work serves as a preliminary exploration into the potential of mixed-reality telepresence in tactical coaching to bridge this gap.

\section{\toolName{} Design}

\subsection{Interviews with Soccer Teams}

Although prior findings have established the foundation for integrating panoramic views or  VR 
into coaching, the challenge of helping players internalize complex tactical ideas persists. To understand the challenges further, we conducted interviews with key personnel from professional clubs. These 45-60 minute interviews were conducted with head coaches, assistant coaches, and performance analysts from City Football Group, Harvard Soccer, and Rice Soccer. The interviews focused on understanding their current challenges in tactical communication, the effectiveness of existing tools, and their expectations from a mixed-reality coaching solution. Three key challenges were identified:
\begin{itemize}
    \item Challenge 1: Difficulty in conveying spatial relationships and team movements from multiple perspectives. 
    \item Challenge 2: Lack of real-time feedback and adjustments during tactical drills, due to voices being unheard on the training ground or players not actually moving while relying on whiteboard visuals. 
    \item Challenge 3: Need for collaborative interaction among players and coaches in a shared environment, due to
players remaining static during whiteboard sessions and coaches struggling to monitor every subtle movement
on the field.
\end{itemize}

\subsection{Prototype Design}
To  address  these  challenges,  we  designed  a  VR  coaching  tool  that  incorporates  the  features  bulleted  below.  For 
development,  we  used  a  Unity  environment  (2022.3.31f1)  with  WebXR  by  De-Panther,  WebGL2,  and  Windows 
builds  for  maximum  compatibility  with  most  personal  devices.  The server infrastructure was supported by Photon PUN2, a third-party networking solution.  
The VR device we used for prototyping is Oculus Quest 2.

\begin{itemize}
    \item \textbf{Multi-perspective viewing across devices}: The tool supports cross-device compatibility (VR, computer, 
smartphone/tablet), enabling remote collaboration between players and coaches. By combining first-person 
view (FPV), a 2D minimap, and broadcast-style video, it enhances spatial understanding and helps players 
better visualize tactical movements, addressing the challenge of conveying complex spatial relationships 
(Challenge 1). 
\item \textbf{Real-time annotation capabilities}: Coaches can provide immediate feedback through both voice and visual 
cues during live sessions, ensuring players actively apply corrections. This feature addresses the challenge of making timely adjustments and overcoming communication barriers during training (Challenge 2). 
\item \textbf{Multiplayer support}: The tool allows coaches and players to interact in real time, with coaches guiding
players via a 2D interface while multiple players engage in a 3D environment. This collaborative setup addresses the challenge of synchronizing team coordination and monitoring player movements more effectively than traditional methods (Challenge 3).
\end{itemize}

\begin{figure*}[t!]
    \includegraphics[width=1.0\textwidth]{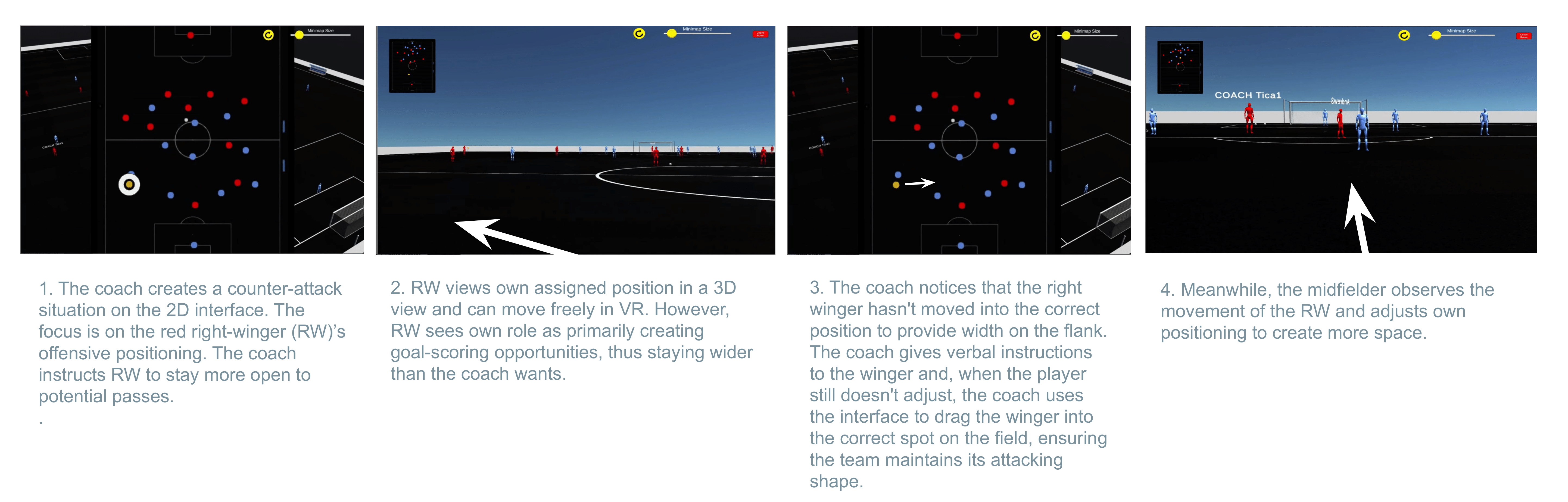}
    \caption{Example \toolName{} use case. Different users are presented with tailored visuals and actions based on their roles.}
    \label{fig:coaching}
\end{figure*}
\section{User Evaluation}

The user evaluation aimed to gather feedback on the initial prototype of \toolName{} and validate its potential to enhance tactical understanding and communication. We conducted feedback sessions with the same key stakeholders from earlier interviewes to assess effectiveness of \toolName{} in meeting the goals of these soccer teams.  During the initial interviews (as shown in Section 3), these organizations expressed the following needs:

\begin{itemize}
    \item \textbf{City Football Group}: Prioritized the need for a user-friendly interface for coaches within sandbox
environments. Envisioned using this tool for self-assessment of retention of tactics in small group training. 
\item \textbf{Harvard Soccer}: Requested features such as adding a ball and drawing 3D arrows for player movement, in 
addition to incorporating short pre-recorded clips for quick tactical meetings. 
\item \textbf{Rice Soccer}: Highlighted the importance of repetitive tactical practice without physical load, specifically for
set plays and strategic shifts during matches. They emphasized separating this tool from physical training
sessions, using it primarily for mental conditioning.
\end{itemize}

During the evaluation, we presented \toolName{} functionality with example tactical coaching scenarios similar to those in Fig.~\ref{fig:coaching} and encouraged coaches to interact with the tool while verbalizing their thoughts on its potential applications. The City Football Group
saw value in the tool for remote learning, especially for loaned-out players (at different clubs) or those recovering
from injury. Harvard appreciated the interactive nature of the tool, noting that it could help maintain player focus during 
meetings by allowing active engagement with the content rather than passively observation of videos. Rice highlighted the 
tool’s effectiveness in reinforcing in-game tactical shifts without adding unnecessary physical strain on the players.

\subsection{Insights}
We summarized the potential benefits of applying \toolName{} to facilitate soccer tactic coaching from user feedback.

\textbf{Rehearsal of Tactical Training Scenarios}: The tool enables repeated practice of specific team movements 
without physical strain, something difficult to achieve in real-world training due to time constraints and player 
fatigue. This allows for focused tactical repetition, leading to better mastery of complex formations and in-
game responses under pressure. 

\textbf{Enhanced Memorability}: The tool strengthens the memory retention of players by immersing them repeatedly in
first-person scenarios, creating vivid mental imprints. This direct engagement is more effective than the 
abstract imagery of tactic boards or passive video analysis. 

\textbf{Attention Guidance}: The interactive platform keeps players actively engaged by letting them interact with the 
coach’s visuals in real-time. Even in a remote environment, coaches can confirm that players understand the
tactics conveyed. This mode is more effective than traditional lecture-style explanations or videos to watch at
home.

\subsection{Limitation and Future Iteration}
Despite  the  positive  feedback  from  the  users,  we  found  several  challenges  in 
deploying  PanoCoach  for  actual  use  in  teams  and  leagues.  First, teams and coaches have varying interests in incorporating an experimental mode of technology into the training workflow of athletes.  As seen in our user evaluation, each team's method of communication tactics varies. For coaches who prefer more traditional or abstract 
communication with whiteboards, a multi-perspective visualization may offer limited benefits. Second, the quality of our visualization
is constrained by the hardware requirements of soccer organizations. More realistic graphics and animation could
create stronger imagery for players to recall during specific game scenarios. Third, the user interface for coaches 
could be improved for greater efficiency. Under the current  implementation, the avatars  must be moved by the 
coach or player. Implementing a rule-based or AI-generated algorithm to automatically generate trajectories could significantly enhance usability and effectiveness for coaches. 

Future iterations of a virtual coaching tool like PanoCoach could focus on refining key features based on feedback 
from  professionals.  In  the  short  term,  effective  improvements might include  automated tracking of
game data from training clips,  expandeded  recording  options,  and  compatibility  with  video  analysis  platforms  commonly used  at  clubs. 
Moreover,  the  benefits  of  this  tool  could  easily  be  expanded  to  other  sports  where synchronized 
movements are crucial, such as basketball or hockey. Our ongoing collaboration with leading sports organizations will also drive this development.
\section{Discussions \& Future Directions}
Our \toolName{} prototype and user evaluation suggested the potential usefulness of mixed reality telepresence in real soccer practices. To realize the entire workflow illustrated in Fig.~\ref{fig:teaser}, there remain several open questions to design and evaluate a multi-person mixed-reality telepresence coaching workflow. Based on insights from this initial study, we point out three challenges and future research directions:

\subsection{Translating movement across views}
The primary objective of integrating mixed-reality telepresence into sports training is to enhance the understanding of team movements, both before and during practice, through effective communication and scenario simulation. 
While \toolName{} allows for flexible transition between 2D and 3D views, these shifts in perspectives can sometimes introduce confusion due to the distinct interaction paradigms. 
Actions that feel intuitive on a 2D touchscreen, such as moving a player from a top-down perspective, may not directly translate to the first-person perspective in an immersive environment.

One critical challenge is maintaining clarity when communicating across the 2D tactical board and the 3D environment. For example, when a coach adjusts a player's position on the 2D tactic board, this movement is immediately apparent. However, in the 3D view, it becomes essential to inform the repositioned player and ensure that other players can anticipate these movements to maintain a cohesive understanding of the movement sequences. 
In addition, the 3D environment enables flexible interactions for players and offers more realistic motion cues in the player avatars, such as the direction a player is facing or the player height and speed—details that might not be as apparent on a 2D board. It is crucial to design interactions and visualizations that effectively convey these nuances across 2D and 3D views to facilitate shared understanding between coaches and players.

To address these challenges, future research can explore the design space for visualizing movement across 2D and 3D co-presence environments. This could involve translating visual cues between 2D and 3D, such as using interpolated or delayed movements of coach-controlled avatars shown in immersive environment, and enhancing interactions between users across views. For example, dynamically displaying and updating a player's position and orientation on both the 2D board and in VR could ensure that coaches and players share a real-time understanding of movement across different perspectives.

\subsection{Visualizing annotation across platforms} 
To facilitate players' understanding of tactics, coaches frequently make annotations on 2D tactical boards. 
Although visualizing 2D annotations to a 3D environment has been extensively studied before, such as placing co-located minimaps or embedding annotations in a 3D environment~\cite{chen2017exploring, willett2016embedded}, 
in physical sports training, augmenting such visualizations onto a 3D view requires additional considerations, particularly from the perspectives of user needs and visualization effectiveness.

First, given the aim of tactical training, it is crucial to determine how much information is useful and necessary to display in VR. Most athletes train to enhance mental and muscle memory, so annotations should not interfere with their physical activity, even during simulated rehearsals in an immersive environment. 
Second, from a design perspective, several limitations need to be addressed when dealing with dynamic spatial data. For example, annotations that are further away may not be easily visible in a standard 3D embedded visualization from a first-person perspective. Additionally, during physical activities, visualizations that need to be interpreted while in motion~\cite{yao2022visualization} should be carefully designed to avoid cognitive overload and ensure they are supportive rather than distracting.

Further research is required to explore how best to convert these annotations into 3D environments to realize the mixed-reality sports training, considering factors such as the optimal amount and type of information to display on each platform, the visibility and accessibility of annotations in a 3D space, and the impact of these visualizations on athletes’ performance during physical activities. 


\subsection{Beyond tactical understanding}
Based on user feedback, rehearsing tactics without engaging in group physical activities has emerged as a promising application of this mixed-reality telepresence training approach. This approach not only facilitates group training with fewer physical constraints but also allows injured players to participate in tactical sessions, maintaining their engagement and understanding of team strategies. As a result, this approach may offer significant potential for innovative training methods in team sports, overcoming both physical and temporal limitations.

However, a significant challenge lies in validating whether these rehearsed tactics are genuinely internalized and can be effectively translated into real-world performance.
One critical issue is the difference between cognitive understanding and physical execution. While players may be able to visualize and mentally rehearse tactics and train their individual movement in a mixed-reality environment, this does not guarantee that they can execute these tactics effectively during actual gameplay with real players. Due to the limitation of visualizations of virtual avatars in AR/VR, the speed, pressure, and environmental factors of a live game may not be properly represented during these simulated training.
Additionally, the lack of physical cues and kinesthetic feedback in a simulated environment can make it difficult for players to fully grasp the timing, spacing, and coordination required for successful tactic execution. For example, in a mixed-reality rehearsal, players might not experience the physical exertion or spatial awareness that comes with actual movement, potentially leading to a gap between their virtual preparation and on-field performance.

To elevate mixed-reality telepresence from tactical understanding to practical training, future research should focus on on integrating key physical data into virtual avatars and environments to more accurately simulate live gameplay. These improvements can help bridge the gap between virtual rehearsals and actual performance, making mixed-reality training more effective. Additionally, longitudinal studies should assess the long-term impact on in-game performance, guiding further directions of mixed-reality telepresence in sports.


\bibliographystyle{abbrv-doi}

\bibliography{template}

\begin{thebibliography}{10}

\bibitem{chen2017exploring}
Z.~Chen, Y.~Wang, T.~Sun, X.~Gao, W.~Chen, Z.~Pan, H.~Qu, and Y.~Wu.
\newblock Exploring the design space of immersive urban analytics.
\newblock {\em Visual Informatics}, 1(2):132--142, 2017.

\bibitem{correia2012}
V.~Correia, D.~Araújo, A.~Cummins, and C.~M. Craig.
\newblock Perceiving and acting upon spaces in a vr rugby task: Expertise effects in affordance detection and task achievement.
\newblock {\em Journal of Sport and Exercise Psychology}, 34(3):305--321, June 2012. doi: {{%
10\hspace{.1pt}\discretionary{.}{%
}{.}\hspace{.4pt}1123\discretionary{/}{%
}{/}jsep\hspace{.1pt}\discretionary{.}{%
}{.}\hspace{.4pt}34\hspace{.1pt}\discretionary{.}{%
}{.}\hspace{.4pt}3\hspace{.1pt}\discretionary{.}{%
}{.}\hspace{.4pt}305}}


\bibitem{faure2019}
C.~Faure, A.~Limballe, B.~Bideau, and R.~Kulpa.
\newblock Virtual reality to assess and train team ball sports performance: A scoping review.
\newblock {\em Journal of Sports Sciences}, 38(2):192--205, November 2019. doi: {{%
10\hspace{.1pt}\discretionary{.}{%
}{.}\hspace{.4pt}1080\discretionary{/}{%
}{/}02640414\hspace{.1pt}\discretionary{.}{%
}{.}\hspace{.4pt}2019\hspace{.1pt}\discretionary{.}{%
}{.}\hspace{.4pt}1689807}}


\bibitem{fortes2021}
L.~S. Fortes et~al.
\newblock Virtual reality promotes greater improvements than video-stimulation screen on perceptual-cognitive skills in young soccer athletes.
\newblock {\em Human Movement Science}, 79:102856, October 2021. doi: {{%
10\hspace{.1pt}\discretionary{.}{%
}{.}\hspace{.4pt}1016\discretionary{/}{%
}{/}j\hspace{.1pt}\discretionary{.}{%
}{.}\hspace{.4pt}humov\hspace{.1pt}\discretionary{.}{%
}{.}\hspace{.4pt}2021\hspace{.1pt}\discretionary{.}{%
}{.}\hspace{.4pt}102856}}


\bibitem{jia2024}
Y.~Jia, X.~Zhou, J.~Yang, and Q.~Fu.
\newblock Animated vr and 360-degree vr to assess and train team sports decision-making: A scoping review.
\newblock {\em Frontiers in Psychology}, 15, July 2024. doi: {{%
10\hspace{.1pt}\discretionary{.}{%
}{.}\hspace{.4pt}3389\discretionary{/}{%
}{/}fpsyg\hspace{.1pt}\discretionary{.}{%
}{.}\hspace{.4pt}2024\hspace{.1pt}\discretionary{.}{%
}{.}\hspace{.4pt}1410132}}


\bibitem{laattala2024}
M.~Laattala, R.~Piitulainen, N.~M. Ady, M.~Tamariz, and P.~Hämäläinen.
\newblock Wave: Anticipatory movement visualization for vr dancing.
\newblock In {\em Proceedings of the CHI Conference on Human Factors in Computing Systems}, May 2024. doi: {{%
10\hspace{.1pt}\discretionary{.}{%
}{.}\hspace{.4pt}1145\discretionary{/}{%
}{/}3613904\hspace{.1pt}\discretionary{.}{%
}{.}\hspace{.4pt}3642145}}


\bibitem{lin2024}
T.~Lin, A.~Aouididi, C.~Zhu-Tian, J.~Beyer, H.~Pfister, and J.-H. Wang.
\newblock Vird: Immersive match video analysis for high-performance badminton coaching.
\newblock {\em IEEE Transactions on Visualization and Computer Graphics}, 30(1):458--468, January 2024. doi: {{%
10\hspace{.1pt}\discretionary{.}{%
}{.}\hspace{.4pt}1109\discretionary{/}{%
}{/}TVCG\hspace{.1pt}\discretionary{.}{%
}{.}\hspace{.4pt}2023\hspace{.1pt}\discretionary{.}{%
}{.}\hspace{.4pt}3327161}}


\bibitem{liu2024}
K.-Y. Liu, T.-Y. Guo, T.-S. Pan, P.-Y. Tung, and Y.-R. Lin.
\newblock Ai batting buddy: A computational and kinematic approach for enhancing batting performance and analysis in baseball.
\newblock In {\em Proceedings of the 2024 International Conference on Multimedia Retrieval}, May 2024. doi: {{%
10\hspace{.1pt}\discretionary{.}{%
}{.}\hspace{.4pt}1145\discretionary{/}{%
}{/}3652583\hspace{.1pt}\discretionary{.}{%
}{.}\hspace{.4pt}3657590}}


\bibitem{ranganathan2007}
R.~Ranganathan and L.~G. Carlton.
\newblock Perception-action coupling and anticipatory performance in baseball batting.
\newblock {\em Journal of Motor Behavior}, 39(5):369--380, September 2007. doi: {{%
10\hspace{.1pt}\discretionary{.}{%
}{.}\hspace{.4pt}3200\discretionary{/}{%
}{/}JMBR\hspace{.1pt}\discretionary{.}{%
}{.}\hspace{.4pt}39\hspace{.1pt}\discretionary{.}{%
}{.}\hspace{.4pt}5\hspace{.1pt}\discretionary{.}{%
}{.}\hspace{.4pt}369\discretionary{%
}{-}{-}380}}


\bibitem{theofilou2022}
G.~Theofilou et~al.
\newblock The effects of a visual stimuli training program on reaction time, cognitive function, and fitness in young soccer players.
\newblock {\em Sensors}, 22(17):6680, September 2022. doi: {{%
10\hspace{.1pt}\discretionary{.}{%
}{.}\hspace{.4pt}3390\discretionary{/}{%
}{/}s22176680}}


\bibitem{tsai2022}
W.-L. Tsai, T.-Y. Pan, and M.-C. Hu.
\newblock Feasibility study on virtual reality based basketball tactic training.
\newblock {\em IEEE Transactions on Visualization and Computer Graphics}, 28(8):2970--2982, August 2022. doi: {{%
10\hspace{.1pt}\discretionary{.}{%
}{.}\hspace{.4pt}1109\discretionary{/}{%
}{/}TVCG\hspace{.1pt}\discretionary{.}{%
}{.}\hspace{.4pt}2020\hspace{.1pt}\discretionary{.}{%
}{.}\hspace{.4pt}3046326}}


\bibitem{vanMaarseveen2018}
M.~J. van Maarseveen, R.~R. Oudejans, D.~L. Mann, and G.~J. Savelsbergh.
\newblock Perceptual-cognitive skill and the in situ performance of soccer players.
\newblock {\em Quarterly Journal of Experimental Psychology}, 71(2):455--470, January 2018. doi: {{%
10\hspace{.1pt}\discretionary{.}{%
}{.}\hspace{.4pt}1080\discretionary{/}{%
}{/}17470218\hspace{.1pt}\discretionary{.}{%
}{.}\hspace{.4pt}2016\hspace{.1pt}\discretionary{.}{%
}{.}\hspace{.4pt}1255236}}


\bibitem{willett2016embedded}
W.~Willett, Y.~Jansen, and P.~Dragicevic.
\newblock Embedded data representations.
\newblock {\em IEEE transactions on visualization and computer graphics}, 23(1):461--470, 2016.

\bibitem{yao2022visualization}
L.~Yao, A.~Bezerianos, R.~Vuillemot, and P.~Isenberg.
\newblock Visualization in motion: A research agenda and two evaluations.
\newblock {\em IEEE Transactions on Visualization and Computer Graphics}, 28(10):3546--3562, 2022.

\end{thebibliography}
\end{document}